\documentclass[aps,prb,floatfix]{revtex4}
\usepackage[]{graphicx}

\usepackage{amsmath}
\usepackage{amsfonts}

\newcommand{\mca}[1]{\mathcal{#1}}
\newcommand{\mr}[1]{\mathrm{#1}}

\newcommand{\bs}[1]{\boldsymbol{#1}}

\newcommand{\fues}[1]{\left(#1\right)}

\newcommand{\yav}[1]{\left[#1\right]}

\newcommand{\llav}[1]{\left\{#1\right\}}

\newcommand{\abs}[1]{\left\vert#1\right\vert}

\newcommand{\rep}[1]{(\ref{#1})}
\newcommand{\pr}{^{\prime}}

\begin{document}
\title{Zero-resistance States Induced by Bichromatic Microwaves}


\author{Alejandro Kunold\footnote{Corresponding
     author: e-mail: {\sf akb@correo.azc.uam.mx}}}
\affiliation{Departamento de Ciencias B\'asicas,
         Universidad Aut\'onoma Metropolitana-Azcapotzalco,
         Av. San Pablo 180,  M\'exico D. F. 02200, M\'exico}
\author{Manuel Torres\footnote{
        e-mail: {\sf torres@fisica.unam.mx}}}
\affiliation{Instituto de F\'{\i}sica,
         Universidad Nacional Aut\'onoma de M\'exico, Apartado Postal
         20-364,  M\'exico Distrito Federal 01000,  M\'exico}
\begin{abstract}
We have studied the bichromatic photoresistance  states of a two dimensional
electron gas in the regime of microwave induced resistance oscillations.
Zudov and coworkers\cite{zudov0} found clear experimental evidence
of zero-resistance states by measuring the bichromatic resistance in
a bidimensional gas of electrons. They found that
the bichromatic resistance closely replicates the superposition
of the two monochromatic components provided that both
contributions are positive. However, the superposition principle
is no longer valid if one of the two contributions give rise to a
zero-resistance state. The experiments by Zudov and coworkers confirm
that negative resistance states are rapidly driven into zero-resistance
states  through A. V. Andreev's symmetry breaking.
In this work we present a  model for the bichromatic-photoconductivity of a
two dimensional electron system subjected  to a uniform magnetic field.
Our model includes both components of the microwave
radiation, a uniform magnetic field and impurity scattering effects.
The conductivity is calculated from a Kubo-like formula.
Our calculations reproduce the main features of Zudov's experimental results.
\end{abstract}

\maketitle                   




\renewcommand{\leftmark}
{A. Kunold, M. Torres: Zero-resistance States by Bichromatic Microwaves}

\section{Introduction.}\label{intro}
Recently, two experimental groups \cite{zudov1,mani1},
reported the  existence  of zero-photoresistance states (ZRS) in ultraclean
two-dimensional electron systems (2DES) subjected to microwave (MW) radiation
and moderate perpendicular magnetic fields.

In these experiments the microwave-induced resistance oscillations
(MIRO) exhibit periods proportional to the inverse of the
magnetic field alternated with ZRS regions
governed by the ratio $\epsilon=  \omega/\omega_c$,
where $\omega_c$ and $\omega$ are the cyclotron an MW frequency.
The   oscillation amplitudes
reach  maxima  at $\epsilon= \omega/\omega_c =j$ and minima at
$\epsilon =j + 1/2$, for $j$ an integer\cite{zudov1}.

Despite the large amount of theoretical work
\cite{ry2,durst,shi,lei,vavilov,dmi1,andre}
the origin of ZRS remains elusive.
Nevertheless most theories agree that some mechanism produces
negative-resistance states (NRS) that rapidly drive the system
into a ZRS due to Andreev's instabilities\cite{andre}.
Some of the models \cite{ry2,durst,vavilov} suggest that the main
cause of NRS is photon-assisted scattering by impurities or disorder.
Alternative explanations \cite{dmi1} propose that NRS
arise from MW induced distribution function fluctuations.
It is important to mention that some theories do not need
to invoke Andreev's instabilities \cite{lee1,platero1}.

In recent experiments \cite{zudov0} the bichromatic resistance
was found to be well described by the superposition of the
two independent monochromatic components in the following terms.
When both monochromatic resistances are positive or zero the bichromatic
resistivity agrees with the sum of both components. If only one
of the monochromatic resistances is zero the bichromatic resistance
is lower than the expected from the direct superposition principle.

In this work we develop a model that is based on the exact
solution of the Shr\"odinger equation of a 2D electron in
the presence of a static magnetic field $B$ interacting with
a MW radiation and a perturbative calculation of randomly
distributed impurities.
For small radiation intensities the conductivity
follows the superposition principle suggested by Zudov \cite{zudov0}. The lowering
of the bichromatic resistivities is shown to originate from
the superposition of NRS and positive resistance  (PR) from the independent
monochromatic contributions.

\section{The Model}
In order to calculate the expectation value of the current density we need the
time dependent density matrix $\rho$ which obeys the von Neumann equation 
$i\partial \rho/\partial t=\yav{H+V_{ext},\rho}$
where $V_{ext}$ is the bias voltage and the Hamiltonian is given by
\begin{equation}
H=H_{\llav{B,\omega_1,\omega_2}}+V\fues{\bs{r}}.
\end{equation}
The MW bichromatic radiation's electric field $\bs{E}=\bs{E}_1+\bs{E}_2=E_x\bs{i}+E_y\bs{j}$  is obtained from
the vector potential$
\bs{A}\fues{t}=-\frac{1}{2}\bs{r}\times\bs{B}+
\mr{Re}\yav{\frac{\bs{E_1}}{\omega_1}\exp\fues{i\omega_1t}
+\frac{\bs{E_2}}{\omega_2}\exp\fues{i\omega_2t}}.
$
Here, the electric fields $\bs{E}_1$ and $\bs{E}_2$ are decomposed in their
polarization vectors $\epsilon_{x1}$, $\epsilon_{y1}$
and $\epsilon_{x2}$, $\epsilon_{y2}$ respectively.
The unperturbed Hamiltonian contains both components of the microwave
radiation and a uniform magnetic field
\begin{eqnarray}
H_{\llav{B,\omega_1,\omega_2}}&=&
\frac{1}{2m^*}\fues{\bs{p}+e\bs{A}}^2+e\bs{E}\cdot \bs x
\nonumber \\
&=&\frac{1}{2m^*}\fues{Q_1^2+P_1^2}
+e E_x\fues{t}\fues{Q_1-P_2}+e E_y\fues{t}\fues{Q_2-P_1},
\end{eqnarray}
where
\begin{eqnarray}
P_1=p_x-\frac{eB}{2}y-\frac{e}{2}\int^tdt\pr E_x\fues{t\pr},\quad\quad
P_2=Q_1-eBx,\\
Q_1=p_y+\frac{eB}{2}x-\frac{e}{2}\int^tdt\pr E_y\fues{t\pr},
\quad\quad Q_2=P_1+eBy 
\end{eqnarray}
are canonical operators that
obey the commutation relations
$\yav{Q_1,P_1}=\yav{Q_2,P_2}=ieB$,
$\yav{Q_1,Q_2}=\yav{P_1,P_2}$ $=\yav{Q_1,P_2}=\yav{Q_2,P_1}=0$.
The impurity scattering potential is given by
\begin{equation}
V\fues{\bs{r}}=\sum_i\int\frac{d^2q}{\fues{2\pi}^2}V\fues{q}
\exp\yav{i\bs{q}\cdot\fues{\bs{r}-\bs{r}_i}}\label{impu}.
\end{equation}
where $m^*$ is the effective mass of the electron,
$\bs{r}_i$ the position of the $i$th impurity and the
explicit form of $V\fues{q}$ depends on the type of impurity.
For the sake of simplicity in this work we will only consider
uncharged impurities.

The unperturbed Hamiltonian $H_{\llav{B,\omega_1,\omega_2}}$
can be reduced to a harmonic oscillator
through a unitary transformation given by\cite{torres1} 
\begin{equation}\label{unit1}
W\fues{t}=\exp\fues{i\eta_1Q_1}\exp\fues{i\zeta_1P_1}
\exp\fues{i\eta_2Q_2}\exp\fues{i\zeta_2P_2}
\exp\fues{-i\int^t dt^{\prime}\mca{L}\fues{t^{\prime}}}
\end{equation}
\begin{figure}
\includegraphics[width=.65\textwidth]{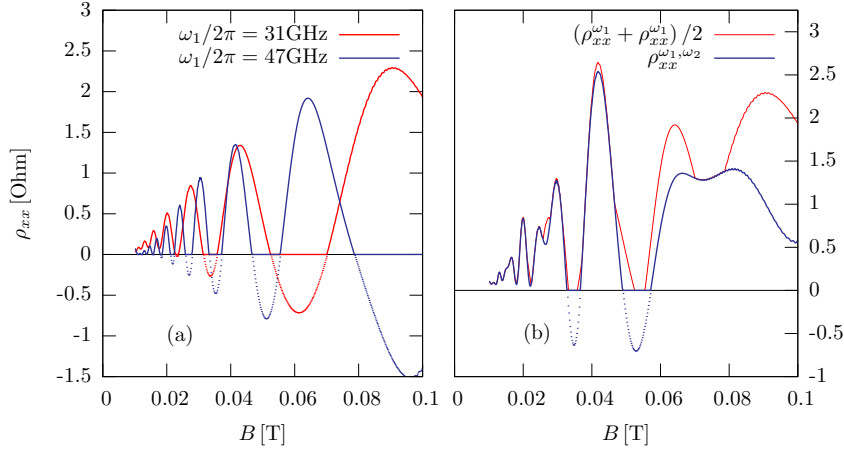}
\caption{Panel (a)  represents the resistivity under
monochromatic radiation for $\omega_1/2\pi=31 {\rm GHz}$
(blue)
and $\omega_2/2\pi=47 {\rm GHz}$ (red). In panel (b) the blue
line shows the bichromatic conductivity $\rho_{xx}$ and
the red line is a plot of $\rho_{xx1}+\rho_{xx2}$.
The dotted lines indicates NRS. In this example
$E_1=220{\rm V/m}$, $E_2=350{\rm V/m}$. The inset
contains plots of successive values of $E_2=350$,
$400$, $450$, $500$, $550{\rm V/m}$.}
\label{fig1}
\end{figure}
where the functions $\eta_i\fues{t}$ and $\zeta_i\fues{t}$ are
solutions of the dynamical equations that follow form the
variation of the classical Lagrangian
$\mca{L}=\frac{\omega_c}{2}\fues{\eta_1^2+\zeta_1^2}+
\dot{\zeta}_1\eta_1+\dot{\zeta}_2\eta_2
+el_B\yav{E_x\fues{\zeta_1+\eta_2}+E_y\fues{\eta_1+\zeta_2}}$
with $l_B=\sqrt{\hbar/eB}$ the magnetic longitude.
The solution to the classical dynamical equations
and the unitary transformation \rep{unit1} constitute the
exact solution for the unperturbed Hamiltonian.
The impurity potential is considered through first order time dependent
perturbation theory.
Solving von Neumann equation in the linear approximation,
the longitudinal conductivity is calculated as
described in \cite{torres1} leading to expressions for
the dark $\sigma_{xx}^D$ and illuminated
$\sigma_{xx}^{\omega_1,\omega_2}$ conductivities. The dark part of the
conductivity leads to Shubnikov-de Haas oscillations.
The illuminated longitudinal conductivity is given by
\begin{equation}\label{conduxx}
\sigma_{xx}^{\omega_1,\omega_2}=
\frac{e^2}{\pi\hbar}\int d\mca{E}\sum_{\mu\nu}\sum_{l_1l_2}
\mr{Im}G_{\mu}\fues{\mca{E}}
B^{l_1l_2}\fues{\mca{E},\mca{E}_\nu}
\abs{\Delta_{\mu\nu}^{l_1l_2\fues{x}}}^2.
\end{equation}
The temperature dependence enters the conductivity through
\begin{equation}
B^{l_1l_2}\fues{\mca{E},\mca{E}_\nu}=-\frac{\partial}{\partial \mca{E}_0}
\llav{\yav{f\fues{\mca{E}+\omega_1l_1+\omega_2l_2+\mca{E}_0}
-f\fues{\mca{E}}}G_{\nu}\fues{\mca{E}+\omega_1l_1+\omega_2l_2+\mca{E}_0}}\
\end{equation}
where $f$ is the Fermi distribution function,
$G_\nu\fues{\mca{E}}=1/\fues{\mca E-i\Gamma}$,
\begin{eqnarray}
\rho_1^{\fues{j}}=\frac{el_BE_1}{2}
\frac{\omega_c\epsilon_{yj}-i\omega_j\epsilon_{xj}}
{\omega_j^2-\omega_c^2+i\omega_j\Gamma}, \quad\quad
\rho_2^{\fues{j}}=\frac{el_BE_2}{2}
\frac{\omega_c\epsilon_{xj}+i\omega_j\epsilon_{yj}}
{\omega_j^2-\omega_c^2+i\omega_j\Gamma}\\
\Delta_i=\frac{e \omega_cl_B^2E_i}
{\omega_i\fues{\omega_i^2-\omega_c^2-i\omega_i\Gamma}}\yav{
\omega_i\fues{q_x\epsilon_{xi}+q_y\epsilon_{yi}}
-i\omega_c\fues{q_x\epsilon_{yi}-q_y\epsilon_{xi}}}
\end{eqnarray}
and $\Gamma=2\pi e /m \mu_0$ is a phenomenological damping factor.
The term
\begin{eqnarray}
\abs{\Delta_{\mu\nu}^{l_1l_2\fues{i}}}^2=
\delta_{\mu\nu}
\yav{\abs{\rho_i^{\fues{1}}}^2\fues{\delta_{l_1,1}
+\delta_{l_1,-1}}\delta_{l_2,0}
+\abs{\rho_i^{\fues{2}}}^2\fues{\delta_{l_2,1}
+\delta_{l_2,-1}}\delta_{l_1,0}}\nonumber\\
+\frac{1}{2}\abs{
\frac{a_iC_{\mu\nu}^{l_1l_2+}}
{\mca{E}_{\mu\nu}+\omega_1l_1+\omega_2l_2-\omega_c}
+\frac{b_iC_{\mu\nu}^{l_1l_2-}}
{\mca{E}_{\mu\nu}+\omega_1l_1+\omega_2l_2+\omega_c}
}
\end{eqnarray}
accounts for the MIRO where $a_x=b_x=1$ and $a_y=-b_y=-i$
and
\begin{eqnarray}
C_{\mu\nu}^{l_1l_2\pm}=
\sqrt{\frac{\mu !}{\nu !}}\fues{\frac{l_B}{\sqrt{2}}}^{\nu-\mu+1}
\sum_i\int\frac{d^2q}{\fues{2\pi}^2}V\fues{q}
e^{-i\bs{q}\cdot\bs{r}_i-\fues{l_Bq}^2/4}
\fues{q_x+iq_y}^{\nu-\mu+1}\fues{\mp q_y-q_x}\nonumber\\
L_\mu^{\nu-\mu}\fues{\frac{l_B^2q^2}{2}}
\fues{\frac{\Delta_1}{\abs{\Delta_1}}}^{l_1}
\fues{\frac{\Delta_2}{\abs{\Delta_2}}}^{l_2}
\yav{J_{l_1}\fues{\abs{\Delta_1}}
J_{l_2}\fues{\abs{\Delta_2}}
-\delta_{l_1,0}\delta_{l_2,0}}\label{lacl1l2}
\end{eqnarray}
with $L_\mu^{\nu-\mu}\fues{l_B^2q^2/2}$ the associated
 Laguerre polynomials.

\section{Superposition Principle}
In most experiments \cite{zudov0}
the intensity of bichromatic radiation is small compared
to the separation of the Landau levels,
that is $l_BeE_{1,2}/\hbar \omega_c \ll 1$,
and the wave length of the microwave radiation is larger
than the magnetic length $l_B$ thus $ql_B\ll 1$ and
$\abs{\Delta_1},\abs{\Delta_2}\ll 1$.
Thus, the two leading terms of the illuminated 
longitudinal conductivity under this conditions
are $\sigma_{xx}^{\omega_1,\omega_2}=\left(\sigma_{xx}^{\omega_1}
+\sigma_{xx}^{\omega_2}\right)/2,$
where $\sigma_{xx}^{\omega_1}$ and $\sigma_{xx}^{\omega_2}$ are
the corresponding monochromatic conductivities given by
\begin{eqnarray}
\sigma_{xx}^{\omega_1}=\frac{e^2}{\pi\hbar}\int d\mca{E}\sum_{\mu\nu}
\mr{Im}G_{\mu}\fues{\mca{E}}
B^{1,0}\fues{\mca{E},\mca{E}_\nu}
\abs{\Delta_{\mu\nu}^{1,0\fues{x}}}^2,\\
\sigma_{xx}^{\omega_2}=
\frac{e^2}{\pi\hbar}\int d\mca{E}\sum_{\mu\nu}
\mr{Im}G_{\mu}\fues{\mca{E}}
B^{0,1}\fues{\mca{E},\mca{E}_\nu}
\abs{\Delta_{\mu\nu}^{0,1\fues{x}}}^2.
\end{eqnarray}
Given the complexity of the previous expressions, the integrals over
the energy where performed numerically.
The bichromatic magnetoresistivity is then
\begin{equation}
\rho^{\omega_1,\omega_2}_{xx}=\frac{\sigma_{xx}^{\omega_1,\omega_2}}
{\fues{\sigma^{\omega_1,\omega_2}_{xx}}^2+\sigma_{xy}^2}
\approx\frac{\sigma^{\omega_1,\omega_2}_{xx}}{\sigma_{xy}^2}
=\frac{\sigma^{\omega_1}_{xx}+\sigma^{\omega_2}_{xx}}{2\sigma_{xy}^2}
\approx \frac{\rho^{\omega_1}_{xx}+\rho^{\omega_2}_{xx}}{2}.
\end{equation}
In Fig. \ref{fig1} we observe the monochromatic components
of the magneto-resistivity [panel (a)]
for frequencies $\omega_1/ 2\pi=31 {\rm GHz} $,
$\omega_2/ 2\pi=47 {\rm GHz} $,
the mobility is $\mu=2\times 10^3 m^2/V s$ and $\Gamma=34.1\times 10^{-6}eV$.
The NRS are plotted with dotted lines.
The bichromatic magneto-resistivity
is plotted in panel (b). We notice that when two positive
resistivities superimpose the bichromatic resistivity is simply
their sum. More specifically, at $B=0.04{\rm T}$ the peaks
from the monochromatic contributions overlap. On the other hand,
when a positive resistivity
is superimposed with a ZRS, the bichromatic resistivity is
suppressed indicating the existence of NRS. At $B=0.06 {\rm T}$,
for example, $\rho_{xx}^{\omega_1}>0$ and $\rho_{xx}^{\omega_2}<0$
thus $\rho_{xx}^{\omega_1,\omega_2}<\rho_{xx}^{\omega_1}
+\rho_{xx}^{\omega_2}$.
In the inset of panel (b) we observe that as $E_2$
arises, it is even possible to produce ZRS (for $E_2=500$ and $550{\rm V/m}$)
 by overlapping
a NRS with a PR state.

\section{Conclusions}
Based on linear response theory we have presented a theoretical
model to explain the MIRO superposition
principle observed by Zudov. We have shown that the interaction of
impurity assisted NRS an PR states from the monochromatic components
are responsible for the enhancement and suppression of
the bichromatic resistivity. 
Our calculations on the bichromatic conductivity
have similar features as those observed by Zudov.

We acknowledge  financial support  by
CONACyT \texttt{J 43110-F}, \texttt{G 32736-E},  and UNAM project No.  \texttt{IN113305}.


\begin{thebibliography}{00}

\bibitem{zudov0}
M.A. Zudov, R.R. Du, L.N. Pfeiffer, and K.W. West
{\em  Phys. Rev. Lett. \/} {\bf 96}, 236804 (2006).

\bibitem{zudov1}
M. A. Zudov, R. R. Du, J. A. Simmons, J. L. Reno 
{\em  Phys. Rev. B \/} {\bf  64}, 201311(R) (2001);
M. A. Zudov, R. R. Du, L. N. Pfeiffer, K. W. West,
{\em  Phys. Rev. Lett \/} {\bf  90}, 046807 (2003);
M. A. Zudov, {\em  Phys. Rev. B  \/}{\bf  69}, 041304(R) (2004).

\bibitem{mani1}
R. G. Mani, J. H. Smet, K. von klitzing, V. Narayanamurti, W.b. Johnson, Umansky,
{\em  Nature \/} {\bf  420}, 646 (2002);
{\em  Phys. Rev. Lett \/} {\bf  92}, 146801 (2004);
R. G. Mani, {\em  Physica E\/} {\bf  22}, 1 (2004).

\bibitem{ry2}
V. I. Ryzhii, R. Suris,
{\em  J Phys. Cond. Matt. \/}{\bf  15}, 6855 (2003).

\bibitem{durst}
A. C. Durst, S. Sachdev, N. Read, S. M. Girvin,
{\em  Phys. Rev. Lett. \/}{\bf  91}, 086803 (2003).

\bibitem{shi}
J. Shi, X. C. Xie,  
{\em  Phys. Rev. Lett. \/}{\bf  91}, 086801 (2003).

\bibitem{lei}
X. L. lei, S. Y. Liu,
{\em  Phys. Rev. Lett. \/}{\bf  91}, 226805 (2003).

\bibitem{vavilov}
M. G. Vavilov, I. L. Aleiner,
{\em  Phys. Rev. B  \/}{\bf  69}, 035303 (2004).

\bibitem{dmi1}
I. A. Dmitriev, A. D. Mirlin, D. G. Polyakov
{\em  Phys. Rev. Lett. \/}{\bf  91}, 226802 (2003);
I. A. Dmitriev, M. G. Vavilov, I. L. Aleiner,  A. D. Mirlin, D. G. Polyakov,
{\em cond-mat/0310668 \/} (2003); 
{\em cond-mat/0409590 \/} (2004).

\bibitem{lee1}D. -H Lee and J. M. Leinaas, Phys. Rev. B {\bf 69}, 115336 (2004).
\bibitem{platero1} J. I\~narrea and G. Platero, Phys. Rev. Lett. {\bf 94}, 016806 (2005).

\bibitem{andre}
A. V. Andreev, I. L. Aleiner, A. J. Millis, 
{\em  Phys. Rev. Lett. \/}{\bf  91}, 056803 (2003).

\bibitem{torres1}
Manuel Torres, Alejandro Kunold
{\em Phys. Rev. B \/}{\bf 71}, 115313 (2005).

\end{thebibliography}
\end{document}